\documentclass[12pt]{article}

\renewcommand{\d}{{\rm d}}
\newcommand{\ve}{\varepsilon}
\begin{document}

\title{Some minisuperspace model for the Faddeev formulation of gravity}

\author{V.M. Khatsymovsky \\
 {\em Budker Institute of Nuclear Physics} \\ {\em of Siberian Branch Russian Academy of Sciences} \\ {\em
 Novosibirsk,
 630090,
 Russia}
\\ {\em E-mail address: khatsym@gmail.com}}
\date{}
\maketitle
\begin{abstract}
We consider Faddeev formulation of general relativity in which the metric is composed of ten vector fields or a $4 \times 10$ tetrad. This formulation reduces to the usual general relativity upon partial use of the field equations.

A distinctive feature of the Faddeev action is its finiteness on the discontinuous fields. This allows to introduce its minisuperspace formulation where the vector fields are constant everywhere on ${\rm I \hspace{-3pt} R}^4$ with exception of a measure zero set (the piecewise constant fields). The fields are parameterized by their constant values {\it independently} chosen in, e. g., the 4-simplices or, say, parallelepipeds into which ${\rm I \hspace{-3pt} R}^4$ can be decomposed. The form of the action for the vector fields of this type is found.

We also consider the piecewise constant vector fields approximating the fixed smooth ones. We check that if the regions in which the vector fields are constant are made arbitrarily small, the minisuperspace action and eqs of motion tend to the continuum Faddeev ones.
\end{abstract}


MSC classes: 83C27

keywords: Einstein theory of gravity; vector fields; minisuperspace model; lattice gravity

\section{Introduction}

Currently discrete methods continue to play an important role in the study of general relativity (GR). Discrete methods may be of interest for numerical calculations; these also provide certain understanding of quantum gravity \cite{Ham'}. In quantum aspect, discretization might be introduced, as in any other field theory, in order to regularize the originally divergent expressions. A distinctive feature of GR from the standard field theoretical viewpoint is its nonrenormalizability. That is, physical amplitudes depend on the specific way of regularization. Consequently, the discrete action must be not only an approximation to the continuum action but the exact continuum action on the field, here metric $g_{\lambda \mu} (x )$, where all except a countable set of the degrees of freedom are frozen, that is, be a minisuperspace theory. An appropriate form of the metric field $g_{\lambda \mu}$ may be that corresponding to the piecewise flat spacetime which can be described as the {\it simplicial complex} composed of the flat 4D tetrahedra or {\it 4-simplices} \cite{piecewiseflat=simplicial'}. The minisuperspace (piecewise flat) spacetime is capable of approximating in some sense any smooth Riemannian manifold spacetime. The usual metric GR on the simplicial complex is known as Regge calculus \cite{Regge'}. For a review of Regge calculus and alternative discrete gravity approaches see, e. g., \cite{RegWil'}. Using Regge calculus as a minisuperspace theory has lead to important results in the Causal Dynamical Triangulations approach to quantum gravity \cite{cdt}.

It is interesting to consider some other minisuperspace formulations of gravity. One of these proceeds from the Faddeev formulation of gravity which represents GR in terms of some extended set of variables which being excluded gives GR.

Faddeev formulation \cite{Fad} is described by $D = 10$ covariant vector fields $f^A_\lambda (x)$. First consider a priori arbitrary $D$. Here, the Latin capitals $A, B, \dots = 1, \dots , D$ refer to an Euclidean (or Minkowsky) $D$-dimensional spacetime, and the Greek indices $\lambda, \mu, \dots = 1, 2, 3, 4$ refer to our four-dimensional spacetime. To simplify notations, we consider the case of the Euclidean metric signature for both the spaces. Our usual metric is a composite field,
\begin{eqnarray}\label{metric}                                              
g_{\lambda \mu} = f^A_\lambda f_{\mu A}.
\end{eqnarray}

\noindent In other words, $f^A_\lambda$ is the $D$-dimensional tetrad.

There are some similarities with the so-called embedded theories of gravity \cite{Deser,Paston}. Our spacetime in these theories is considered as a four-dimensional hypersurface with some coordinates $f^A (x^1, x^2, x^3, x^4)$ in a flat $D$-dimensional spacetime. This corresponds to the choice $f^A_\lambda = \partial_\lambda f^A$. The number $D$ of the functions $f^A$ is sufficient to parameterize all ten independent components of the metric (\ref{metric}) at $D \geq 10$. Contrary to that, $f^A_\lambda$ are freely chosen in the Faddeev formulation.

More generally, the Faddeev gravity has to do with gravity theories where the metric spacetime is not a fundamental physical concept but emerges from a non-spatio-temporal structure present in a more complete theory of interacting fundamental constituents (appearing, e. g., in the context of string theory) \cite{Chiu}.

An important point of the Faddeev approach is introducing a connection
\begin{equation}                                                            
\tilde{\Gamma}_{\lambda \mu\nu} = f^A_\lambda f_{\mu A, \nu} ~~~ (f_{\mu A, \nu} \equiv \partial_\nu f_{\mu A}), ~~~ \tilde{\Gamma}^\lambda_{\mu\nu} = g^{\lambda\rho} \tilde{\Gamma}_{\rho \mu \nu}
\end{equation}

\noindent alternative to the unique torsion-free Levi-Civita one, $\Gamma^\lambda_{ \mu\nu}$. We can introduce the projectors onto the {\it vertical} $\Pi_{AB}$ and {\it horizontal} $\Pi_{||AB}$ directions,
\begin{equation}\label{Pi, Pi||}                                            
\Pi_{AB} = \delta_{AB} - f^\lambda_A f_{\lambda B}, ~~~ \Pi_{||AB} = f^\lambda_A f_{\lambda B}.
\end{equation}

\noindent Then the curvature tensor reads
\begin{equation}                                                            
\hspace{-15mm} K^\lambda_{\mu \nu \rho} = \tilde{\Gamma}^\lambda_{\mu\rho, \nu} - \tilde{\Gamma}^\lambda_{\mu\nu, \rho} + \tilde{\Gamma}^\lambda_{\sigma\nu} \tilde{\Gamma}^\sigma_{\mu\rho} - \tilde{\Gamma}^\lambda_{\sigma\rho} \tilde{\Gamma}^\sigma_{\mu\nu}
= \Pi^{AB} (f^\lambda_{A, \nu} f_{\mu B, \rho} - f^\lambda_{A, \rho} f_{\mu B, \nu}).
\end{equation}

\noindent Note that the projector $\Pi_{AB}$ makes the usual derivatives equivalent to the covariant ones,
\begin{equation}                                                            
\Pi^{AB} f_{\lambda B, \mu} = \Pi^{AB} \nabla_\mu f_{\lambda B}.
\end{equation}

\noindent The action takes the form
\begin{equation}\label{Fad action}                                          
\hspace{-10mm} S = \int {\cal L} \d^4 x = \int K^\lambda_{\mu \lambda \rho} g^{\mu \rho} \sqrt {g} \d^4 x = \int \Pi^{AB} (f^\lambda_{A, \lambda} f^\mu_{B, \mu} - f^\lambda_{A, \mu} f^\mu_{B, \lambda}) \sqrt {g} \d^4 x.
\end{equation}

Varying the action gives the equations of motion,
\begin{equation}\label{delta-S/delta-f}                                     
\frac{\delta S}{2 \sqrt{g} \delta f^\lambda_A} = f^A_\nu \left ( K^\nu_\lambda - \frac{1}{2} \delta^\nu_\lambda K \right ) + \Pi^{AB} \left ( f^\nu_{B, \nu} T^\mu_{\lambda \mu} + f^\nu_{B, \mu} T^\mu_{\nu \lambda} + f^\nu_{B, \lambda} T^\mu_{\mu \nu} \right ) = 0
\end{equation}

\noindent where $K^\nu_\lambda = g^{\mu \rho} K^\nu_{\mu \lambda \rho}$, $K = K^\lambda_\lambda$, and $T^\lambda_{ \mu \nu } = f^{\lambda A} (f_{\mu A , \nu} - f_{\nu A , \mu})$ is torsion. Projecting (\ref{delta-S/delta-f}) by $\Pi_{AB}$ leaves the second term in the LHS,
\begin{equation}\label{V lambda A}                                          
b^\nu{}_{\nu A} T^\mu_{\lambda \mu} + b^\nu{}_{\mu A} T^\mu_{\nu \lambda} + b^\nu{}_{\lambda A} T^\mu_{\mu \nu} = 0.
\end{equation}

\noindent Here, $b^\lambda{}_{\mu A} = \Pi_{AB} f^{\lambda B}_{, \mu} \equiv \Pi_{AB} \nabla_\mu f^{\lambda B}$. The index $A$ of the projected by $\Pi_{AB}$ expression takes on effectively $D - 4$ values. Thus we have $4 (D - 4)$ independent equations. If these equations are considered as a linear system for $T^\lambda_{\mu \nu}$, the number of these equations might be sufficient to ensure that $4 \times 6 = 24$ components of $T^\lambda_{\mu \nu}$ be zero just at $D \geq 10$.

From now on, we set $D = 10$. In this case, the matrix of the system (\ref{V lambda A}) is a square $24 \times 24$ matrix (a function of $b^\lambda{}_{\mu A}$), and the determinant of it is nonzero for random values of $b^\lambda{}_{\mu A}$ \cite{Fad}. Therefore, it is assumed that the vertical equations of motion are equivalent to vanishing torsion: $T^\lambda_{\mu \nu} = 0$. This means that the connection is the unique torsion-free Levi-Civita one, $\tilde{\Gamma}^\lambda_{\mu\nu} = \Gamma^\lambda_{\mu\nu}$, and the curvature tensor is Riemannian one, $K^\lambda_{\mu \nu \rho} = R^\lambda_{\mu \nu \rho}$. Then the first term in the LHS of (\ref{delta-S/delta-f}) just gives the Einstein equations.

In the present paper, we study a minisuperspace formulation of the Faddeev gravity on the continuum fields $f^A_\lambda (x )$ where all except a countable set of the degrees of freedom are frozen.

Regge calculus can be considered as GR on the stepwise metric field, but only under the condition that the induced on any hyperplane metric should be continuous while crossing this hyperplane, that is, the different 4-simplices should match on their common faces. In the Faddeev gravity, the vector fields can be discontinuous, that is, their values can be taken absolutely independently in the different 4-simplices. This point is considered in more detail in Section \ref{piecewise}. In that Section, we directly evaluate the Faddeev action on the piecewise constant fields $f^A_\lambda (x )$ on the 4-simplices or, more generally, some polytopes. Our estimate is close in spirit to the work by Friedberg and Lee \cite{Fried} where the Regge action has been derived in the usual GR by estimating the Hilbert-Einstein action on some metric describing the piecewise flat manifold. In Section \ref{cont-limit}, we show that if our minisuperspace stepwise form of $f^A_\lambda (x )$ is chosen to approximate some fixed smooth $f^A_\lambda (x )$ arbitrarily closely, the discrete Faddeev action (and eqs of motion) tend to the continuum one. This calculation is close in spirit to the work by Feinberg, Friedberg, Lee, and Ren \cite{Fein} where the Regge action has been shown to tend to the Hilbert-Einstein one if simplicial decomposition of the given smooth manifold is made finer and finer. Our calculation is made for some simple case of decomposing spacetime into polytopes, namely, decomposing into cuboids. In the usual GR, the requirement that the cuboids should match on their common faces leads to that the manifold composed of the flat cuboids is flat as well. In the Faddeev gravity, the possibility for the fields $f^A_\lambda (x )$ being discontinuous makes it possible to use the flat cuboids with independent $f^A_\lambda (x )$ for modeling the curved spacetimes. Some application of this possibility is discussed in Conclusion where also the paper is summarized and some possible further problem is posed.

\section{Faddeev action on the piecewise constant fields}\label{piecewise}

For clarity, it is helpful to illustrate our consideration by an example of the usual GR on the same footing. Suppose $n$ is any of the numbers 1, 2, 3, 4, and $\alpha, \beta, \gamma, ... $ run over these numbers with exception of $n$. Suppose the interval takes the form $\d s^2 = g_{nn} (\d x^n)^2 + g_{\alpha \beta} \d x^\alpha \d x^\beta$, $g_{n \alpha } = 0$ (an analog of gauge fixing by choosing the {\it stationary} coordinate frame). In GR, the derivatives over $x^n$ squared in the Hilbert-Einstein action are
\begin{equation}                                                            
\int R \sqrt{g} \d^4 x = \frac{1}{4 } \int (g^{\alpha\gamma}g^{\beta\delta} - g^{\alpha\beta}g^{\gamma\delta})g_{\alpha\beta , n}g_{\gamma\delta , n}g^{nn}\sqrt{g}\d^4x + ...~.
\end{equation}

\noindent Such form of the action implies that $g_{\alpha\beta}$ is at least continuous everywhere as a function of $x^n$,
\begin{equation}\label{delta-g}                                            
g_{\alpha \beta}|_{x^n = x^n_0 + 0} - g_{\alpha \beta}|_{x^n = x^n_0 - 0} = 0
\end{equation}

\noindent at any $x^n_0$. Relaxing this requirement leads to infinity in the action and is therefore prohibited in GR. In other words, the metric induced on the hypersurface $x^n = x^n_0$ should be a continuous function of $x^n_0$. We can consider some particular ansatz for the metric field in which ${\rm I \hspace{-3pt} R}^4$ as the set of points $x = (x^1, x^2, x^3, x^4)$ is divided by the hypersurfaces $a_\lambda x^\lambda + b = 0$ (mathematical hyperplanes) into polytopes, in particular, 4-simplices. Then the metric is introduced and assumed to be constant in each of these 4-simplices. Thus we have the spacetime decomposed into the flat 4-simplices. Analogously to the paper \cite{Fried}, this metric distribution could be substituted into the Einstein action to yield the Regge action. Some condition of the type of (\ref{delta-g}) states that the metric induced on any 3-face from each of the two adjacent 4-simplices should be the same. In other words, the 4-simplices should match on their common faces. At the same time, the normal metric $g_{nn}$ {\it should} be discontinuous on some 3-faces, otherwise the full metric $g_{\lambda \mu}$ could be extended as a constant to the whole spacetime resulting in the flat spacetime.

Let us pass to the Faddeev gravity. Again, we can divide ${\rm I \hspace{-3pt} R}^4$ by the hypersurfaces $a_\lambda x^\lambda + b = 0$ into polytopes like the 4-simplices or parallelepipeds and take $f^\lambda_A (x )$ to be constant in each polytope. Now there is no the square of any derivative in the action (\ref{Fad action}), and its existence does not imply any conditions for $f^\lambda_A (x )$ on the 3-faces like the condition (\ref{delta-g}) for $g_{\lambda \mu}$. That is, $f^A_\lambda (x )$ (and thus $g_{\lambda \mu} = f^A_\lambda f_{\mu A}$) are independent in the different polytopes (if the eqs of motion are not taken into account). Of course, this does not mean that these fields do not propagate: the action (\ref{d f d f V Pi dx}) below describes an interaction of the fields in the neighboring 4-simplices. The action on this field distribution turns out to be the sum of contributions from the triangles forming the 2-dimensional polytope's faces whose evaluation is the same for the different types of polytopes. Therefore, in this Section we consider the case of the 4-simplices as the most universal one. The field $f^\lambda_A$ in the most part of some neighborhood of any 3-simplex $\sigma^3$ can be taken to depend (in the stepwise manner) only on one coordinate. Therefore, the contribution to $S$ from $\sigma^3$ is zero. A contribution to $S$ comes from an arbitrarily small neighborhood of any 2-simplex $\sigma^2$ due to a dependence on some two coordinates, say, $x^1$, $x^2$. (We assume that $x^1$, $x^2$ are constant on $\sigma^2$.) Evidently, the expression $(f^\lambda_{A, \lambda} f^\mu_{B, \mu} - f^\lambda_{A, \mu} f^\mu_{B, \lambda})$ appearing in $S$ has support on $\sigma^2$. That is, it is the $\delta$-function $const \cdot \delta (x^1) \delta (x^2)$. To fix the constant, we take into account the fact that this expression is a full derivative,
\begin{equation}                                                           
f^\lambda_{A, \lambda} f^\mu_{B, \mu} - f^\mu_{A, \lambda} f^\lambda_{B, \mu} = \partial_\lambda Q^\lambda, ~~~ Q^\lambda = f^\lambda_A \partial_\mu f^\mu_B - f^\mu_A \partial_\mu f^\lambda_B.
\end{equation}

\noindent Then the integral over any neighborhood of the point $(x^1, x^2) = (0, 0)$ (which defines this constant) reduces to a contour integral not depending on the details of the behavior of the fields at this point. We have
\begin{equation}\label{oint_C}                                             
\hspace{-20mm} \int (f^\lambda_{A, \lambda} f^\mu_{B, \mu} - f^\lambda_{A, \mu} f^\mu_{B, \lambda}) \d x^1 \d x^2 = \oint_C (f^1_A \d f^2_B - f^2_A \d  f^1_B).
\end{equation}

\begin{figure}[h]
\unitlength 1pt
\begin{picture}(200,200)(-200,-100)
\put(0,0){\line(-1,0){100}}
\put(0,0){\line(0,-1){95}}
\put(0,0){\vector(1,0){100}}
\put(0,0){\vector(0,1){95}}

\put(0,0){\circle{40}}
\put(16,12){\vector(-1,1){4}}
\put(83,31){$\sigma^3_1$}
\put(41,82){$\sigma^3_2$}
\put(5,90){$x^2$}
\put(105,-3){$x^1$}
\put(-49,65){$\sigma^3_3$}
\put(-100,37){$\sigma^3_{i-1}$}
\put(-82,-65){$\sigma^3_i$}
\put(-25,-91){$\sigma^3_{i+1}$}
\put(61,-65){$\sigma^3_{n-1}$}
\put(83,-17){$\sigma^3_n$}
\put(5,1){$\sigma^2$}
\put(-11,-10){$O$}
\put(17,15){$C$}
\put(40,40){$\sigma^4_2$}
\put(55,7){$\sigma^4_1$}
\put(-15,50){$\sigma^4_3$}
\put(-53,35){\dots}
\put(-60,-13){$\sigma^4_i$}
\put(-40,-50){$\sigma^4_{i+1}$}
\put(10,-50){\dots}
\put(47,-28){$\sigma^4_n$}

\thicklines

\put(17.5,7){\line(5,2){62.5}}
\put(0,0){\line(1,2){40}}
\put(0,0){\line(-2,3){40}}
\put(0,0){\line(-5,2){80}}
\put(-12.5,-10){\line(-5,-4){57.5}}
\put(0,0){\line(-1,-5){16}}
\put(0,0){\line(6,-1){80}}
\put(0,0){\line(1,-1){60}}

\end{picture}

\caption{Some neighborhood of a triangle $\sigma^2$ shared by 3- and 4-simplices.}
\label{sigma2}
\end{figure}
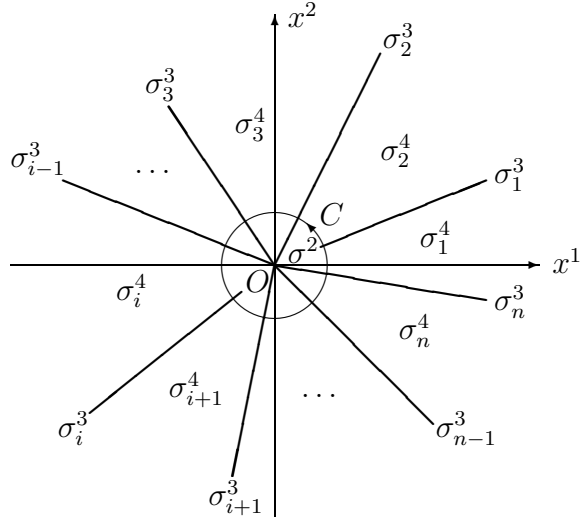

\noindent In fig. \ref{sigma2}, the center $O$ which represents the 2-simplex $\sigma^2$ is encircled by the integration contour $C$ counterclockwise.

There are some products of step functions and delta functions under the contour integral sign in (\ref{oint_C}) which can be defined ambiguously depending on the intermediate regularization. Formally, we can write $\theta (x) \delta (x) = \theta (0) \delta (x)$ where we can take for $\theta (0)$ any number $\alpha$ from the interval $[0,1]$. Assuming the geometry of fig. \ref{sigma2}, this choice means the equation
\begin{equation}\label{f-sigma3}                                           
f(\sigma^3_i) = (1 - \alpha ) f(\sigma^4_i ) + \alpha f(\sigma^4_{i+1} )
\end{equation}

\noindent for the values of a function $f$ on the 3-face $\sigma^3_i$ in terms of its values on the two 4-simplices $\sigma^4_i$ and $\sigma^4_{i+1}$ sharing this 3-face. Here, $f$ is $f^\lambda_A$ or $f^\lambda_B$, $\lambda = 1,2$. As a result, the symmetrized over $A, B$ integral (\ref{oint_C}) is
\begin{eqnarray}\label{delta f delta f}                                    
& & \oint_C (f^1_A \d f^2_B - f^2_A \d  f^1_B) + (A \leftrightarrow B) \nonumber \\
& & = \sum^n_{i = 1} \left \{ f^1_A (\sigma^3_i) [ f^2_B (\sigma^4_{i+1} ) - f^2_B (\sigma^4_i ) ] - f^2_B (\sigma^3_i) [ f^1_A (\sigma^4_{i+1} ) - f^1_A (\sigma^4_i ) ] \right \} + (A \leftrightarrow B) \nonumber \\ & & = \sum^n_{i=1} [ f^1_A ( \sigma^4_i ) f^2_B ( \sigma^4_{i+1} ) - f^1_A ( \sigma^4_{i+1} ) f^2_B ( \sigma^4_i ) ] + (A \leftrightarrow B).
\end{eqnarray}

\noindent Here we use (\ref{f-sigma3}), and the dependence on $\alpha$ remarkably disappears. Thus,
\begin{eqnarray}\label{d f d f}                                            
& & 2 ( f^\lambda_{A, \lambda} f^\mu_{B, \mu} - f^\lambda_{A, \mu} f^\mu_{B, \lambda} ) \nonumber\\ & & = \delta (x^1 ) \delta (x^2 ) \sum^n_{i=1} [ f^1_A ( \sigma^4_i ) f^2_B ( \sigma^4_{i+1} ) - f^1_A ( \sigma^4_{i+1} ) f^2_B ( \sigma^4_i ) ] + (A \leftrightarrow B).
\end{eqnarray}

Next, we would like to multiply (\ref{d f d f}) by $\Pi^{AB} \sqrt{g}$. Since the latter function is not unambiguous at $(x^1, x^2) \to (0, 0)$, this product can not be defined unambiguously. We can only write
\begin{eqnarray}\label{d f d f V Pi}                                       
(f^\lambda_{A, \lambda} f^\mu_{B, \mu} - f^\lambda_{A, \mu} f^\mu_{B, \lambda}) \tilde{\Pi}^{AB} & = & \delta (x^1 ) \delta (x^2 ) \tilde{\Pi}^{AB} (\sigma^2 ) \sum^n_{i=1} \left [ f^1_A ( \sigma^4_i ) f^2_B ( \sigma^4_{i+1} ) \right. \nonumber \\ & & \phantom{ \delta (x^1 ) \delta (x^2 ) \tilde{\Pi}^{AB} (\sigma^2 ) \sum^n_{i=1} } \left. - f^1_A ( \sigma^4_{i+1} ) f^2_B ( \sigma^4_i ) \right ].
\end{eqnarray}

\noindent Here, $\tilde{\Pi}^{AB} = \Pi^{AB} \sqrt{g}$, and $\tilde{\Pi}^{AB} (\sigma^2 )$ is some effective value of $\tilde{\Pi}^{AB}$ on $\sigma^2$. It specifies (or, more correctly, is specified by) the details of the intermediate regularization of $f^\lambda_A ( x )$ used in a neighborhood of the points of discontinuity $(x^1, x^2, x^3, x^4) = (0, 0, x^3, x^4)$ forming the triangle $\sigma^2$. Some natural choice for $\tilde{\Pi}^{AB} (\sigma^2 )$ might be some average of $\tilde{\Pi}^{AB} (\sigma^4_i )$, $\sigma^4_i \supset \sigma^2$, see, e. g., (\ref{tilde-Pi-sigma2=average-Pi-sigma4}) below.

Suppose that the triangle $\sigma^2$ is spanned by some edges $\sigma^1_3, \sigma^1_4$ having some 4-vectors $\Delta x^\lambda_{\sigma^1_3}$, $\Delta x^\lambda_{\sigma^1_4}$. Our choice of the coordinates $x^\lambda$ above ($x^1, x^2$ are constant on $\sigma^2$) assumes that $\Delta x^1_{\sigma^1_3} = \Delta x^2_{\sigma^1_3} = \Delta x^1_{\sigma^1_4} = \Delta x^2_{\sigma^1_4} = 0$. Integrating (\ref{d f d f V Pi}) in the neighborhood of $\sigma^2$ we find the contribution of $\sigma^2$ to the action to be
\begin{eqnarray}\label{d f d f V Pi dx}                                    
\int (f^\lambda_{A, \lambda} f^\mu_{B, \mu} - f^\lambda_{A, \mu} f^\mu_{B, \lambda}) \tilde{\Pi}^{AB} d^4 x & = & \frac{1}{2} \tilde{\Pi}^{AB} (\sigma^2 ) \sum^n_{i=1} \left [ f^1_A ( \sigma^4_i ) f^2_B ( \sigma^4_{i+1} ) \right. \nonumber \\ & & \hspace{-15mm} \left. - f^1_A ( \sigma^4_{i+1} ) f^2_B ( \sigma^4_i ) \right ] ( \Delta x^3_{\sigma^1_3} \Delta x^4_{\sigma^1_4} - \Delta x^3_{\sigma^1_4} \Delta x^4_{\sigma^1_3} ).
\end{eqnarray}

This can be rewritten in an invariant form. It is convenient to decompose the world vectors into the components $f^{\sigma^1_\lambda}_A$ associated with the two chosen edges $\sigma^1_3$, $\sigma^1_4$ and another two $\sigma^1_1$, $\sigma^1_2$ such that $\sigma^1_1$, $\sigma^1_2$, $\sigma^1_3$, $\sigma^1_4$ span some $\sigma^4_{i_0} \supset \sigma^2$,
\begin{equation}                                                           
f^\lambda_A (\sigma^4_i ) = \sum_\mu f^{\sigma^1_\mu}_A (\sigma^4_i ) \Delta x^\lambda_{\sigma^1_\mu}.
\end{equation}

\noindent The action (\ref{d f d f V Pi dx}) reads
\begin{equation}\label{df df Pi V det Dx}                                  
\hspace{-0mm} \frac{1}{2} \Pi^{AB} (\sigma^2 ) \sqrt{g (\sigma^2 )} \det \| \Delta x^\lambda_{\sigma^1_\mu} \| \sum^n_{i=1} \left [ f^{\sigma^1_1}_A ( \sigma^4_i ) f^{\sigma^1_2}_B ( \sigma^4_{i+1} ) - f^{\sigma^1_1}_A ( \sigma^4_{i+1} ) f^{\sigma^1_2}_B ( \sigma^4_i ) \right ].
\end{equation}

\noindent Here, $\Pi^A_B = \delta^A_B - \sum_\lambda f^A_{\sigma^1_\lambda } f_B^{\sigma^1_\lambda }$, $f^A_{\sigma^1_\lambda } = f^A_\mu \Delta x^\mu_{\sigma^1_\lambda }$, $\sqrt{g } \det \| \Delta x^\lambda_{\sigma^1_\mu} \| = \sqrt{ \det \| f^A_{\sigma^1_\lambda } f_{ \sigma^1_\mu A }\|}$ (a volume-type value). The physical variables $f^{\sigma^1_\lambda}_A$ )\footnote{It is more appropriate to associate, say, $f^{\sigma^1_1}_A$ not with the edge $\sigma^1_1$, but with the tetrahedron spanned by $\sigma^1_2$, $\sigma^1_3$, $\sigma^1_4$ or rather with the vertex in $\sigma^4_{i_0}$ opposite to this tetrahedron.} or $f_{\sigma^1_\lambda}^A$ are world invariants, and the action (\ref{df df Pi V det Dx}) itself is explicitly a world invariant too.

\section{Approximating fixed smooth field $f^\lambda_A (x)$ distribution}\label{cont-limit}

Suppose that the considered stepwise form of $f^\lambda_A (x )$ should be chosen to approximate some fixed smooth $f^\lambda_A (x )$ arbitrarily closely. To this end, we can assign to $f^\lambda_A (\sigma^4 )$ the value of $f^\lambda_A (x )$ at some point $x = x_{\sigma^4 }$: $f^\lambda_A (\sigma^4 ) = f^\lambda_A (x_{\sigma^4 } )$. We could define $x_{\sigma^4 }$ to be equidistant to the (five) vertices of $\sigma^4$ (the dual Voronoi lattice vertex).

To study possible effectiveness of the stepwise approximation, consider certain simplifying {\it modification of the simplicial} decomposition. Namely, consider the case in which ${\rm I \hspace{-3pt} R}^4$ as the set of points $x = (x^1, x^2, x^3, x^4)$ is divided not into the 4-simplices, but into the rectangular parallelepipeds or cuboids
\begin{equation}                                                           
\{ x | x^\lambda_0 + n^\lambda \ve^\lambda \leq x^\lambda \leq x^\lambda_0 + (n^\lambda + 1) \ve^\lambda, \lambda = 1, 2, 3, 4 \}
\end{equation}

\noindent numbered by four integers $(n^1, n^2, n^3, n^4)$. Then the field $f^\lambda_A (x )$ is introduced and assumed to be constant in each of these cuboids.

Let $\sigma^k$ denotes not simplices but cuboids (only in this Section). The above derivation leading to (\ref{d f d f V Pi dx}) remains the same. Only the RHS of (\ref{d f d f V Pi dx}) is multiplied by 2, and $n = 4$ there. The product of the $\delta$-function (\ref{d f d f}) and step function $\tilde{\Pi}^{AB} (x )$ in (\ref{d f d f V Pi}) should be defined by using some intermediate regularization, e. g., the point splitting one. That is, we can take the product of $(f^\lambda_{A, \lambda} f^\mu_{B, \mu} - f^\lambda_{A, \mu} f^\mu_{B, \lambda}) (x )$ and $\tilde{\Pi}^{AB} (x + \delta )$ and then put $\delta \to 0$. In this way, we obtain $\tilde{\Pi}^{AB} (\sigma^2 ) = \tilde{\Pi}^{AB} (\sigma^4_{i_0})$ in (\ref{d f d f V Pi dx}) where $\sigma^4_{i_0}$ is one of the four cuboids sharing the rectangle $\sigma^2$. We can average the action obtained over the signs of the components $\delta^1, \delta^2$. Thus
\begin{equation}\label{tilde-Pi-sigma2=average-Pi-sigma4}                  
\tilde{\Pi}^{AB} (\sigma^2 ) = \frac{1}{4} \sum^4_{i = 1} \tilde{\Pi}^{AB} (\sigma^4_{i}).
\end{equation}

\begin{figure}[h]
\unitlength 1pt
\begin{picture}(160,120)(-220,-50)
\put(0,0){\line(-1,0){90}}
\put(0,0){\line(0,-1){55}}
\put(0,0){\vector(1,0){80}}
\put(0,0){\vector(0,1){65}}

\put(60,40){\line(-1,0){120}}
\put(60,-40){\line(-1,0){120}}
\put(60,40){\line(0,-1){80}}
\put(-60,40){\line(0,-1){80}}

\put(30,20){\circle*{2}}
\put(-30,20){\circle*{2}}
\put(30,-20){\circle*{2}}
\put(-30,-20){\circle*{2}}

\put(5,60){$x^2$}
\put(85,-3){$x^1$}

\put(45,45){$(\ve^1 , \ve^2 )$}
\put(-85,45){$(-\ve^1 , \ve^2 )$}
\put(-85,-51){$(-\ve^1 , -\ve^2 )$}
\put(45,-51){$(\ve^1 , -\ve^2 )$}
\put(2,4){$(0 , 0 )$}

\put(10,25){${\scriptstyle (\ve^1 / 2 , \ve^2 / 2 )}$}
\put(-57,25){${\scriptstyle (-\ve^1 / 2 , \ve^2 / 2 )}$}
\put(-57,-28){${\scriptstyle (-\ve^1 / 2 , -\ve^2 / 2 )}$}
\put(10,-28){${\scriptstyle (\ve^1 / 2 , -\ve^2 / 2 )}$}

\end{picture}

\caption{The four 4-cuboids sharing a quadrangle at $x^1 = 0, x^2 = 0$.}
\label{quadrangle-action}
\end{figure}
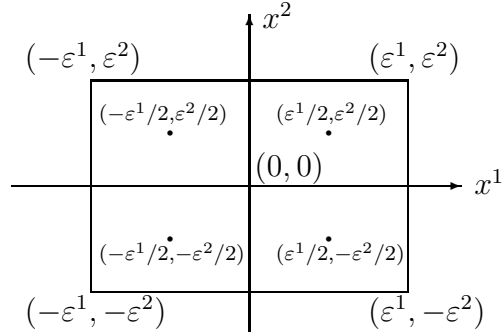

Given a smooth $f^\lambda_A (x )$, the value of $f^\lambda_A (x_{\sigma^4 } )$ is defined as $f^\lambda_A (x )$ at the center $x = x_{\sigma^4 }$ of the cuboid $\sigma^4$. Consider the contribution of some quadrangle $\sigma^2$ in the $x^3, x^4$-plane whose coordinates $(x^1, x^2)$ are taken to be $(0, 0)$ (fig. \ref{quadrangle-action}).

\noindent We have
\begin{eqnarray}                                                           
\tilde{\Pi}^{AB} ( \sigma^2 ) = \frac{1}{4} \left [ \tilde{\Pi}^{AB} \left ( \frac{\ve^1}{2}, \frac{\ve^2}{2} \right ) + \tilde{\Pi}^{AB} \left ( - \frac{\ve^1}{2}, \frac{\ve^2}{2} \right ) + \tilde{\Pi}^{AB} \left ( - \frac{\ve^1}{2}, - \frac{\ve^2}{2} \right ) \right. \nonumber \\ \left.  + \tilde{\Pi}^{AB} \left ( \frac{\ve^1}{2}, - \frac{\ve^2}{2} \right ) \right ] = \tilde{\Pi}^{AB} ( 0, 0) + O([\ve ]^2)
\end{eqnarray}

\noindent where $[\ve ]$ is some typical discretization scale, say, the maximum number of $\ve^1, \ve^2, \ve^3, \ve^4$. Besides that,
\begin{eqnarray}                                                           
& & \sum^4_{i=1} [ f^1_A ( \sigma^4_i ) f^2_B ( \sigma^4_{i+1} ) - f^1_A ( \sigma^4_{i+1} ) f^2_B ( \sigma^4_i ) ] ~ = ~
\sum^4_{i=1} f^1_A ( \sigma^4_i ) [ f^2_B ( \sigma^4_{i+1} ) - f^2_B ( \sigma^4_{i-1} ) ] \nonumber \\ & & = ~~~ f^1_A \left ( \frac{\ve^1}{2}, \frac{\ve^2}{2} \right ) \left [ f^2_B \left ( - \frac{\ve^1}{2}, \frac{\ve^2}{2} \right ) - f^2_B \left ( \frac{\ve^1}{2}, - \frac{\ve^2}{2} \right ) \right ] \nonumber \\ & & ~ + ~ f^1_A \left ( - \frac{\ve^1}{2}, \frac{\ve^2}{2} \right ) \left [ f^2_B \left ( - \frac{\ve^1}{2}, - \frac{\ve^2}{2} \right ) - f^2_B \left ( \frac{\ve^1}{2}, \frac{\ve^2}{2} \right ) \right ] \nonumber \\ & & ~ + ~ f^1_A \left ( - \frac{\ve^1}{2}, - \frac{\ve^2}{2} \right ) \left [ f^2_B \left ( \frac{\ve^1}{2}, - \frac{\ve^2}{2} \right ) - f^2_B \left ( - \frac{\ve^1}{2}, \frac{\ve^2}{2} \right ) \right ] \nonumber \\ & & ~ + ~ f^1_A \left ( \frac{\ve^1}{2}, - \frac{\ve^2}{2} \right ) \left [ f^2_B \left ( \frac{\ve^1}{2}, \frac{\ve^2}{2} \right ) - f^2_B \left ( - \frac{\ve^1}{2}, - \frac{\ve^2}{2} \right ) \right ] \nonumber
\end{eqnarray}
\begin{eqnarray}                                                           
& & \hspace{-5mm} = ~~ \left [ f^1_A \left ( \frac{\ve^1}{2}, \frac{\ve^2}{2} \right ) - f^1_A \left ( - \frac{\ve^1}{2}, - \frac{\ve^2}{2} \right ) \right ] \left [ f^2_B \left ( - \frac{\ve^1}{2}, \frac{\ve^2}{2} \right ) - f^2_B \left ( \frac{\ve^1}{2}, - \frac{\ve^2}{2} \right ) \right ] \nonumber \\ & & + ~ \left [ f^1_A \left ( - \frac{\ve^1}{2}, \frac{\ve^2}{2} \right ) - f^1_A \left ( \frac{\ve^1}{2}, - \frac{\ve^2}{2} \right ) \right ] \left [ f^2_B \left ( - \frac{\ve^1}{2}, - \frac{\ve^2}{2} \right ) - f^2_B \left ( \frac{\ve^1}{2}, \frac{\ve^2}{2} \right ) \right ] \nonumber
\end{eqnarray}
\begin{eqnarray}                                                           
& & \hspace{-13mm} = ~~ [\ve^1 f^1_{A, 1} + \ve^2 f^1_{A, 2} + O([\ve ]^3) ] [ - \ve^1 f^2_{B, 1} + \ve^2 f^2_{B, 2} + O([\ve ]^3) ] \nonumber \\ & & \hspace{-11mm} + ~ [ - \ve^1 f^1_{A, 1} + \ve^2 f^1_{A, 2} + O([\ve ]^3) ] [ - \ve^1 f^2_{B, 1} - \ve^2 f^2_{B, 2} + O([\ve ]^3) ] \nonumber
\end{eqnarray}
\begin{eqnarray}                                                           
& & \hspace{-13mm} = ~~ 2 \ve^1 \ve^2 (f^1_{A, 1} f^2_{B, 2} - f^1_{A, 2} f^2_{B, 1} + O([\ve ]^4) \nonumber
\end{eqnarray}
\vspace{-9mm}
\begin{eqnarray}\label{ff-ff-cube}                                         
& & \hspace{-13mm} = ~~ \ve^1 \ve^2 \sum^2_{\lambda , \mu = 1} (f^\lambda_{A, \lambda } f^\mu_{B, \mu } - f^\lambda_{A, \mu } f^\mu_{B, \lambda }) + O([\ve ]^4).
\end{eqnarray}

\noindent Only the arguments $x^1, x^2$ of the functions are shown. The arguments $x^3, x^4$ are those of the center of some quadrangle, e. g., $\ve^3 / 2, \ve^4 / 2$. The integral of $\delta (x^1 ) \delta (x^2 )$ over $\d^4 x$ in some neighborhood of $\sigma^2$ is
\begin{equation}                                                           
\Delta x^3_{\sigma^1_3} \Delta x^4_{\sigma^1_4} - \Delta x^3_{\sigma^1_4} \Delta x^4_{\sigma^1_3} = \ve^3 \ve^4.
\end{equation}

\noindent In total, the contribution of the quadrangle in the $x^3, x^4$ plane is
\begin{eqnarray}\label{Pi-ff-DxDx-fin-dif}                                 
& & \tilde{\Pi}^{AB} (\sigma^2 ) \sum^4_{i=1} \left [ f^1_A ( \sigma^4_i ) f^2_B ( \sigma^4_{i+1} ) - f^1_A ( \sigma^4_{i+1} ) f^2_B ( \sigma^4_i ) \right ] ( \Delta x^3_{\sigma^1_3} \Delta x^4_{\sigma^1_4} - \Delta x^3_{\sigma^1_4} \Delta x^4_{\sigma^1_3} ) \nonumber \\ & & \!\! = ~~ \ve^1 \ve^2 \ve^3 \ve^4 \tilde{\Pi}^{AB} \sum^2_{\lambda , \mu = 1} (f^\lambda_{A, \lambda } f^\mu_{B, \mu } - f^\lambda_{A, \mu } f^\mu_{B, \lambda }) + O([\ve ]^6).
\end{eqnarray}

\noindent The arguments are $(x^1, x^2, x^3, x^4 ) = (0, 0, \ve^3 /2, \ve^4 /2 )$, and these can be shifted by $(n^1, n^2, n^3, n^4 )$. Then (\ref{Pi-ff-DxDx-fin-dif}) can be summed over $(n^1, n^2, n^3, n^4 )$. Thus we obtain up to $O([\ve ]^2)$ the bulk contribution to the finite difference formula for some continuum integral expression differing from the Faddeev action (\ref{Fad action}) only by that the summation over $\lambda, \mu$ is performed only over $\lambda, \mu = 1, 2$. But we should take into account also the contributions of the quadrangles located in the other than $x^3, x^4$ five 2-planes $x^\nu, x^\rho$. This will give us just the Faddeev action (\ref{Fad action}).

\begin{figure}[h]
\unitlength 1pt
\begin{picture}(160,120)(-220,-50)
\put(0,0){\line(-1,0){90}}
\put(0,0){\line(0,-1){55}}
\put(0,0){\vector(1,0){80}}
\put(0,0){\vector(0,1){65}}

\put(60,45){\line(-1,0){120}}
\put(60,-45){\line(-1,0){120}}
\put(60,45){\line(0,-1){90}}
\put(-60,45){\line(0,-1){90}}
\put(60,15){\line(-1,0){120}}
\put(60,-15){\line(-1,0){120}}
\put(20,45){\line(0,-1){90}}
\put(-20,45){\line(0,-1){90}}

\put(40,30){\circle*{2}}
\put(-40,30){\circle*{2}}
\put(40,-30){\circle*{2}}
\put(-40,-30){\circle*{2}}
\put(40,0){\circle*{2}}
\put(0,0){\circle*{2}}
\put(0,30){\circle*{2}}
\put(-40,0){\circle*{2}}
\put(0,-30){\circle*{2}}

\put(20,15){\circle{3}}
\put(-20,15){\circle{3}}
\put(-20,-15){\circle{3}}
\put(20,-15){\circle{3}}

\put(5,60){$x^2$}
\put(85,-3){$x^1$}

\put(36,33){${\scriptstyle ( \ve^{\! 1} \! , \ve^{\! 2} \! )}$}
\put(-59,33){${\scriptstyle ( \!\! -\ve^{\! 1} \! , \ve^{\! 2} \! )}$}
\put(-59,-38){${\scriptstyle ( \!\! -\ve^{\! 1} \! , \! -\ve^{\! 2} \! )}$}
\put(35,-38){${\scriptstyle ( \ve^{\! 1} \! , \! -\ve^{\! 2} \! )}$}
\put(1,3){${\scriptstyle (0 \! , 0 )}$}
\put(36,3){${\scriptstyle ( \ve^{\! 1} \! , 0 \! )}$}
\put(1,33){${\scriptstyle ( 0 \! , \ve^{\! 2} \! )}$}
\put(-59,3){${\scriptstyle ( \!\! -\ve^{\! 1} \! , 0 \! )}$}
\put(0,-38){${\scriptstyle ( \! 0 \! , \! - \! \ve^{\! 2} \! )}$}

\put(20,18){${\scriptstyle ( \ve^{\! 1} \!\! / \! 2 \! , \ve^{\! 2} \!\! / \! 2 \! )}$}
\put(-52,18){${\scriptstyle ( \!\! -\ve^{\! 1} \!\! / \! 2 \! , \ve^{\! 2} \!\! / \! 2 \! )}$}
\put(-57,-23){${\scriptstyle ( \!\! -\ve^{\! 1} \!\! / \! 2 \! , \! -\ve^{\! 2} \!\! / \! 2 \! )}$}
\put(20,-23){${\scriptstyle ( \ve^{\! 1} \!\! / \! 2 \! , \! -\ve^{\! 2} \!\! / \! 2 \! )}$}

\end{picture}

\caption{To the eqs of motion $\partial S / \partial f^\lambda_A (\sigma^4 ) = 0$, $\sigma^4$ being centered at $x = 0$.}
\label{quadrangle-eqs-of-motion}
\end{figure}
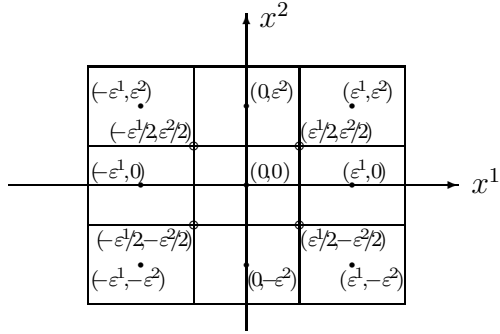

As far as the eqs of motion $\partial S / \partial f^\lambda_A (\sigma^4 ) = 0$ are concerned, the contributions from those quadrangles which are the 2-faces of $\sigma^4$ take part in forming these equations. As a result, these eqs depend on $f^\mu_B (\sigma^{4 \prime } )$ where the cuboids $\sigma^{4 \prime }$ have common 2-faces with $\sigma^4$. It is convenient to consider separately the contributions to $S$ from any four quadrangles of $\sigma^4$ with the same orientation, say, those located in the $x^3, x^4$-planes, see fig. \ref{quadrangle-eqs-of-motion}.

\noindent The bilinears $f^1_A ( \sigma^4 ) f^2_B ( \sigma^{4 \prime } )$ in the discrete action arise for those cuboids $\sigma^{4 \prime }$ which have common 3-faces with $\sigma^4$. In fig. \ref{quadrangle-eqs-of-motion}, $\sigma^4$ is depicted as the square centered at $(x^1, x^2 ) = (0, 0)$. Then $\sigma^{4 \prime }$ correspond to the squares centered at $( \ve^1 , 0)$, $( 0, \ve^2 )$, $( - \ve^1 , 0)$, and $( 0, - \ve^2 )$. As earlier, the discrete variables $f^\lambda_A (\sigma^4 )$ are assumed to be the values of some smooth field $f^\lambda_A (x )$ at the centers of cuboids $x_{\sigma^4}$, $f^\lambda_A (\sigma^4 ) = f^\lambda_A (x_{\sigma^4} )$. In these notations, e. g., the bilinear $f^1_A (0, 0) f^2_B (0, \ve^2 )$ (only the arguments $x^1, x^2$ are shown) arises with opposite signs in the contributions of the quadrangles at $( - \ve^1 / 2, \ve^2 / 2 )$ and at $( \ve^1 / 2, \ve^2 / 2 )$ being multiplied there by $\tilde{\Pi}^{AB} ( - \ve^1 / 2, \ve^2 / 2 )$ and $\tilde{\Pi}^{AB} ( \ve^1 / 2, \ve^2 / 2 )$, respectively. Thus, we obtain the terms in the eqs of motion $\partial S / \partial f^\lambda_A (0 ) = 0$ containing the difference of $\tilde{\Pi}^{AB}$ at the close points $\tilde{\Pi}^{AB} ( - \ve^1 / 2, \ve^2 / 2 ) - \tilde{\Pi}^{AB} ( \ve^1 / 2, \ve^2 / 2 )$. In the continuum version, this corresponds to the derivative $\partial_1 \tilde{\Pi}^{AB}$.

Another origin of the derivatives of $\tilde{\Pi}^{AB}$ in the eqs of motion is varying $\tilde{\Pi}^{AB}$ w. r. t. $f^\lambda_C$. In the continuum version, these two contributions are related via
\begin{equation}                                                           
\tilde{\Pi}^{AB}_{, \lambda} = \frac{\partial \tilde{\Pi}^{AB}}{\partial f^\nu_C} f^\nu_{C, \lambda}.
\end{equation}

\noindent This provides matching different terms in the eqs of motion just leading to the required form of these eqs. (\ref{delta-S/delta-f}), in particular, the vertical ones (\ref{V lambda A}) which result in zero torsion $T^\lambda_{\mu \nu}$ and classical equivalence to the Einstein general relativity. In the discrete case, we still have the exact derivative $\partial \tilde{\Pi}^{AB} / \partial f^\nu_C$ due to varying $\tilde{\Pi}^{AB}$ over $f^\lambda_C$, but the above finite difference of $\tilde{\Pi}^{AB}$ at some close points arises instead of the derivative when varying the bilinear in $f^\lambda_A$ factor in the action. Then the different terms in the eqs of motion generally mismatch as compared to the continuum case. That is, the eqs of motion for our minisuperspace system differs from the naive finite-difference version of the continuum eqs (\ref{delta-S/delta-f}) by nonzero RHS having its structure distinct from LHS and vanishing at $[\ve ] \to 0$. In particular, the finite-difference version of torsion $T^\lambda_{\mu \nu}$ vanishes (and classical equivalence to the discrete GR, Regge calculus, takes place) not identically, but at $[\ve ] \to 0$.

\section{Conclusion}

A peculiar feature of the minisuperspace Faddeev formulation of gravity is that the values of the fields can be chosen independently in the different 4-simplices. This is an advantage as compared to the Regge calculus where the condition of continuity of the metric induced on the 3-faces of the type of (\ref{delta-g}) is required. This does not mean that the fields $f^A_\lambda$ do not propagate: the action (\ref{d f d f V Pi dx}) describes an interaction of the fields in the neighboring 4-simplices. This off shell (that is, without taking into account the eqs of motion, virtual) independence of the fields $f^A_\lambda$ means independence of the metric $g_{\lambda \mu} = f^A_\lambda f_{\mu A}$ and thus independence of (the edge lengths of) the 4-simplices.

An application of the lack of need for any continuity condition for $f^A_\lambda (x )$ is the possibility to use flat cuboids with {\it independent} constant $f^A_\lambda (x )$ for modeling the curved spacetimes. The minisuperspace cuboid action looks combinatorially simple in comparison with the simplicial one and looks like the continuum action where the derivatives are replaced by some finite-difference approximation (see, e. g., (\ref{ff-ff-cube})).

An application can be to any surface as a set of the virtually {\it independent} triangles, and this gives us a new approach to the quantum problem of definition of the spectrum of surface area as the sum of spectra of separate triangles.

To summarize, an analog of the Regge calculus is obtained with smaller geometrical restrictions on its building blocks, polytopes. The set of physical variables consists of the world invariants $f^A_{\sigma^1} (\sigma^4 )$ or $f_A^{\sigma^1} (\sigma^4 )$ (the edge 10-vectors), independent for the different 4-simplices $\sigma^4$ even for the same edge $\sigma^1$. We have evaluated the Faddeev action on the piecewise constant fields $f^A_\lambda (x )$. Besides that, we have considered the obtained minisuperspace action and eqs of motion in the limit when the piecewise constant $f^A_\lambda$ serve to approximate some fixed smooth $f^A_\lambda (x )$, and the sizes of the regions of constancy, 4-cuboids, tend to zero (the continuum limit). We have found that the action and the eqs of motion tend to the continuum ones in the continuum limit. The classical equivalence to the usual GR (Regge calculus in the discrete case) generally takes place approximately, up to the terms vanishing in the continuum limit.

The minisuperspace Faddeev action obtained depends on the intermediate regularization of the fields in some neighborhood of the two-dimensional skeleton of the simplicial complex (2-simplices). This dependence is, however, inessential for correct reproducing the continuum Faddeev action in the continuum limit. Therefore, the following question arises to be further studied: can we specify the regularized behavior of the minisuperspace fields $f^A_\lambda (x )$ in the vicinity of the 3-dimensional simplicial boundary so that the minisuperspace Faddeev action be {\it exactly} equivalent to the Regge action on classical level?

\section*{Acknowledgments}

The author thanks I.A. Taimanov who had attracted the author's attention to the new formulation of gravity and Ya.V. Bazaikin for valuable discussions on this subject. The author is grateful to I.B. Khriplovich who has provided moral support, A.A. Pomeransky and A.S.Rudenko for discussion at a seminar which has stimulated writing this article. The present work was supported by the Ministry of Education and Science of the Russian Federation.

\end{document}